\documentstyle[pra,preprint,eqsecnum,aps,tighten]{revtex}
\input epsf
\begin{document}

\tighten

\def\itp#1#2{\hfill{NSF-ITP-{#1}-{#2}}}
\def\cs{{\cal S}}
\def\tr{{\rm Tr}}

%\today

\title{Entropy of Classical Histories}

\author{{\bf Todd A. Brun}\thanks{Current address: Physics Department,
Carnegie Mellon University, Pittsburgh, PA  15213-3890.
Email:  tbrun@andrew.cmu.edu} \\
{\sl Institute for Theoretical Physics, \\
University of California, Santa Barbara,
CA 93106-4030} \\
and\\
{\bf James B.~Hartle} \\
{\sl Department of Physics, \\
 University of California, Santa Barbara, CA  931069530}}

\maketitle

\begin{abstract}
We consider a number of proposals for the entropy of sets of classical
coarse-grained
histories based on the procedures of Jaynes, and prove a series
of inequalities relating these measures.  We then examine these as
a function of the coarse-graining for various classical systems,
and show explicitly
that the entropy is minimized by the finest-grained
description of a set of histories.  
We propose an extension of the  second
law of thermodynamics to the entropy of histories. We briefly
discuss the implications for decoherent or consistent
 history formulations of quantum mechanics.
\end{abstract}


\vfil

\itp{97}{102}\\

\maketitle

\setlength{\baselineskip}{.3in}
\setcounter{footnote}{0}

\section{Introduction}
\label{sec: I}

Entropies are measures of the information missing from a
coarse-grained description of a system. Different coarse-grained 
descriptions give rise to different entropies. If an entropy
is low at one time, it will have a general tendency to grow as  
its coarse-graining is translated forward in time. 
That is the second law of
thermodynamics.

Usually, entropy is constructed from a coarse-grained description
at a single moment in
time. For example, if all that is known of the state of a system at a
particular time is its total energy, the missing information is the
entropy of the microcanonical ensemble --- a quantity which is
independent of time. If all that is known at a time are the
expected values of the energy, number, and momentum densities, averaged over
volumes large enough to be in local equilibrium, then the missing information
per volume is the time-dependent entropy density of hydrodynamics. 

The Jaynes procedure \cite{KZ,Ros83}
gives a general method for constructing the entropy
of a system at a moment of time. To illustrate, let $M$ be the phase space of
a classical system and $\rho(x), ~x \in M$  be the probability distribution
representing the state of the system. 
Suppose $A(x)$ is a classical 
quantity whose expected
value $\langle A \rangle$ is known, where

\begin{equation}
\langle A \rangle ~=~ \int_M ~dx ~A(x) ~\rho(x)\ .
\label{oneone}
\end{equation}
The missing information $S$ is
constructed by maximizing the entropy functional

\begin{equation}
\cs (\tilde{\rho})~ =~ - \int_M ~dx ~\tilde\rho(x) ~\log_2 
~\left[\tilde\rho(x) \right]
\label{onetwo}
\end{equation}
over all $\tilde\rho (x)$ that imply the same expected value.
In symbols,

\begin{equation}
S ~=~ \max_{\tilde{\rho}} ~\cs(\tilde{\rho}) \mid_{\langle A
\rangle_{\tilde{\rho}}~=~ \langle A \rangle_\rho}\ .
\label{onethree}
\end{equation}

However, entropy need not only apply to coarse-grained alternatives
at one moment in time. More generally, one can consider the missing information
of a sequence of alternatives at a succession
of times. These are the entropies of
coarse-grained {\it histories} of the system. A variety of such entropies 
have been described ({\it e.g.}\cite{LP88,GH90a,IL97})
and applied to measures of coarse-graining\cite{GH90a},
classicality\cite{GH95}, and effective complexity\cite{GL96}.
In theories which possess a notion of history but lack a fixed notion of
time (such as certain formulations of quantum gravity \cite{Har95c})
 the missing information of histories may be the only notion of entropy
available.

In this paper we examine the entropy of histories for classical stochastic
systems --- classical systems with a probabilistic law of evolution (including
deterministic evolution as a special case). We use Isham's history space
\cite{Ish94,IL94} and a
generalization of the Jaynes procedure to give a unified view of several
different kinds of entropies for histories and describe relations among
them. We illustrate with numerical
calculations in some simple examples.
Finally, we describe a modest generalization of the second law of
thermodynamics applicable to
the entropy of histories and test it in a simple model. Our considerations are
almost
entirely classical, but in Section V we
point the way to generalizations for the quantum
mechanical case.

\pagebreak
\section{Entropies of Histories}

\subsection{Histories and History Space}

We consider classical theories with (most generally) a stochastic evolution
law in a space $M$
through a finest-grained net of $N$ times
separated by equal intervals $\eta$.
For this discussion the space $M$ could be a configuration space of
particle positions, a spatial lattice, or a phase space.
We denote a point in $M$ by $x$.

The cases when $M$ is a discrete space or a continuous manifold differ only
formally, and can to a large extent be treated together by using a common
notation. We define

\begin{mathletters}
\label{twoone}
\begin{eqnarray}
\tr \, f(x) &=&\sum_{x \in M} ~f(x)\label{twoone a}
\end{eqnarray}
when $M$ is  discrete, and 
\begin{eqnarray}
\tr \, f(x) &=& \int_M dx ~f(x)\label{twoone b}
\end{eqnarray}
when M is continuous. This notation is suggestive for the quantum
mechanical case to be treated later. We also define
\begin{eqnarray}
V &\equiv&\tr(I)\ ,
\label{twoone c}
\end{eqnarray}
\end{mathletters}

\noindent
where $I$ is the unit function on $M$. $V$ 
is an integer when $M$ is discrete and a real number when $M$
is continuous.

A fine-grained history is described by a sequence of
$x_A$, $A=1$, $\cdots$, $N$
for each of the finest-grained net of times. Histories are therefore naturally
thought of as living in a classical ``history space" ${\bf M} ~=~ M \times
~\cdots~\times M$,
 with one factor for each fine-grained time. A point in ${\bf
M}$ is denoted by ${\bf x}$, and corresponds to a fine-grained history. This is
the classical analog of the ``history space" introduced by Isham \cite{Ish94},
and used so effectively in quantum theory by Isham and Linden \cite{IL94,IL97},
and Isham, Linden and Schreckenberg \cite{ILS94}.

A coarse-grained\footnote{The notion of coarse-graining has many 
specific applications in physics. An anonymous referee suggested
\cite{GasXX} as a convenient reference to some of these.}  set of alternative histories is a partition of the set
${\bf M}$ of fine-grained
histories into an exhaustive set of mutually exclusive regions
or classes
$c_\alpha$. Each class is a single {\it coarse-grained history}. 
We can usefully introduce projections onto these regions of ${\bf M}$,

\begin{equation}
{\bf P}_\alpha ({\bf x}) = \cases{1 & ${\bf x} \in c_\alpha\ ,$ \cr 
0 &${\bf x}
\notin c_\alpha\ .$\cr}
\label{twotwo}
\end{equation}

A sequence of coarse-grained
alternatives at a series of times $t_1$, $\cdots$, $t_n$
is an example of a coarse-grained history. Suppose the
alternatives at time $t_k$ are whether $x$ is in one of a set of regions of $M$,
$\left\{\Delta^k_{\alpha_k}\right\},\, \alpha_k = 1, 2, \dots$,
with volumes $V^k_{\alpha_k}$. We 
introduce projections on these regions of $M$,

\begin{equation} \label{twothree}
P^k_{\alpha_k} (x) = \cases{1 & $x \in \Delta^k_{\alpha_k}\ ,$ \cr 
0 &$x \notin \Delta^k_{\alpha_k}\ ,$\cr} 
\end{equation} 
which satisfy

\begin{equation} \label{twoA}
P^k_{\alpha_k} (x) ~P^k_{\alpha'_k} (x)
  ~=~ \delta_{\alpha_k \alpha'_k} ~ P^k_{\alpha_k} (x)
\end{equation} 
and

\begin{equation} \label{twoB}
\tr \left(P^k_{\alpha_k} \right) ~=~ V^k_{\alpha_k}\ .
\end{equation} 
In this case, a coarse-grained
 history is a particular sequence of regions $\alpha ~\equiv~(\alpha_n, \ldots,
\alpha_1)$ and corresponds to a projection on ${\bf M}$ of the form

\begin{equation} \label{twofour}
{\bf P}_\alpha = I \times \cdots \times P^n_{\alpha_n} \times I \times
\cdots \times P^1_{\alpha_1} \times \cdots  \times I\ . 
\end{equation} 
That is, ${\bf P}_\alpha$ is
the projection on ${\bf M}$
with projections $P^k_{a_k}$ inserted at the times $t_k$ and $I$'s
at all  other times. In the discrete case, the most general projection 
${\bf
P}_\alpha$ can always be written as a sum of such chains:

\begin{equation} \label{twofive}
{\bf P}_\alpha ({\bf x}) ~=~ \sum_{\alpha_1 ~\cdots~\alpha_n \in \alpha} ~ {\bf
P}_{\alpha_n ~\cdots~\alpha_1} ({\bf x})\ ,
\end{equation} 
which allows the construction of a {\it narrative}
for each coarse-grained history. For histories of the form (\ref{twofour})
it would read:
``the system was in $\Delta^1_{\alpha_1}$ at $t_1$,
then $\Delta^2_{\alpha_2}$ at $t_2$, \ldots'' In the
continuum case there is a corresponding integral.

We assume that there is a probability law for the fine-grained histories, that
is, a probability function ${\bf W}({\bf x})$
on ${\bf M}$. ${\bf W}({\bf x})$ satisfies

\begin{equation} \label{twosix}
{\bf W} ({\bf x}) \geq 0\ ,  \quad {\rm and}\quad \tr ({\bf W}) ~=~ 1\ .
\end{equation} 
Of course, ${\bf W}({\bf x})$ may have special forms
in particular circumstances. For example,
for a Markov process

\begin{equation} \label{twoseven}
{\bf W} ({\bf x}) ~=~ p_\eta(x_N \mid x_{N-1}) ~p_\eta(x_{N-1} \mid x_{N-2})
\cdots ~p_\eta(x_2 \mid x_1) ~\rho(x_0)\ ,
\end{equation} 
where $\rho(x_0)$ is the distribution at the initial time and
$p_\eta(x | y)$
is the transition probability to arrive at $x$ in a time $\eta$ having 
started from $y$. 
If $M$ were a classical phase space, deterministic evolution would
be represented by (\ref{twoseven}) with

\begin{equation} \label{twoeight}
p_\eta (x | y) ~=~ \delta(x - x_\eta (y))\ ,
\end{equation} 
where $x_\eta (y)$ is the phase-space point $y$ evolved by the time $\eta$.

The probability of a coarse-grained history is 

\begin{equation} \label{twonine}
p_\alpha =  \tr \left({\bf P}_\alpha {\bf W}\right)\ .
\end{equation} 
For example, in the case of a sequence of
alternatives like (\ref{twothree}) at
a series of times, and a Markovian probability of the form (\ref{twoseven}),

\begin{eqnarray} \label{twoten}
p_{\alpha_n \cdots \alpha_1} &=~ &\int dx_n ~\cdots
  \int dx_1 ~ P^n_{\alpha_n} (x_n)
~p(x_n t_n ~|~ x_{n-1} t_{n-1}) ~P^{n-1}_{\alpha_{n-1}} (x_{n-1}) \nonumber \\
& & \cdots P^1_{\alpha_1}(x_1)  ~p(x_1 t_1 ~|~x_0 t_0) ~\rho(x_0)\nonumber
\\ 
& &\equiv \int\ dx_0\, C_{\alpha_n\cdots\alpha_1} (x_0)\, \rho\, (x_0)\ .
\end{eqnarray} 
Here $p(x' t' ~|~ xt)$ is the composition of all the $p_\eta$'s from $t$ to
$t'$, $t_0$ is the initial time, and $C_{\alpha_n\cdots\alpha_1} (x_0)$ is
defined to be the probability of a coarse-grained history
$\alpha_1\cdots\alpha_n$  given that the
system is initially at $x_0$.

\subsection{The Entropy of Histories}

The Jaynes construction may now be applied in history space to give an entropy
for histories. We introduce the entropy 
functional\footnote[1]{In the continuous
case, where ${\bf W}$ is probability density, rather than a probability, eqn.
(\ref{twoeleven}) is not generally invariant under dimensional transformations.
However, (2.13) is a standard definition. Dimensionally invariant quantities may
be obtained by appropriate subtractions, e.g. $- \log_2 ~V^N$, or better, by
rescaling the coordinates so they are dimensionless.}

\begin{equation} \label{twoeleven}
\cs({\bf W}) ~=~ - \tr ({\bf W}~ \log_2~ {\bf W})\ .
\end{equation} 
(Unable to introduce a bold-faced calligraphic S, we rely on the argument 
of $\cal S$ to
distinguish this definition from (\ref{onetwo})).

The {\it history space entropy}
 $S_{hs}(\{c_\alpha\})$ of a set of coarse-grained alternative
histories is then 
\begin{equation} \label{twotwelve}
S_{hs} (\{c_\alpha\}) \equiv
  \max_{\widetilde{{\bf W}}} \cs ({\widetilde {\bf W}}) 
\mid_{\tr({\bf P}_\alpha {\widetilde {\bf W}}) = 
\tr({\bf P}_\alpha {\bf W})}\ .
\end{equation} 
In words, $S_{hs}$ maximizes the missing information $\cs$ over all probability
distributions $\widetilde{{\bf W}}$ on ${\bf M}$ that reproduce the 
probabilities of the
coarse-grained histories $\{c_\alpha\}$ following from ${\bf W}$.

The important property of $S_{hs}(\{c_\alpha\})$ is that it increases on
coarse-graining. Specifically, suppose 
$\{\overline{c}_{\overline{\alpha}}\}$ is a
coarse-graining of the set $\{c_\alpha\}$. That means that
$\{\overline{c}_{\overline{\alpha}}\}$ is a partition of the 
$\{c_\alpha\}$ into
larger classes, and 

\begin{equation} \label{twothirteen}
\overline{c}_{\overline{\alpha}} = \mathop{\cup}_{\alpha \in \overline{\alpha}}
 c_\alpha \ .
\end{equation} 
Then, as with any Jaynes type construction,

\begin{equation} \label{twofourteen}
S_{hs} (\{c_\alpha\}) \leq S_{hs}(\{\overline{c}_{\overline{\alpha}}\}\ .
\end{equation} 
The proof is immediate from (\ref{twotwelve}). The constraints for 
$\{c_\alpha\}$
contain those for $\{\overline{c}_{\overline{\alpha}}\}$ but there are more of
them. The maximum therefore can only be less.

Since the ${\bf P_\alpha}({\bf x})$ are mutually exclusive projections, an
expression for $S_{hs}$ can be derived by carrying out the maximization using
Lagrange multipliers to enforce the constraints. The result is

\begin{equation} \label{twofifteen}
S_{hs}(\{c_\alpha\}) = - \sum_\alpha p_\alpha \log_2 p_\alpha + \sum_\alpha 
p_\alpha \log_2  \tr({\bf P}_\alpha)\ .
\end{equation} 
The maximum value of $S_{hs}$ occurs for the coarsest-grained set of 
histories set
where the only history with nonzero
probability is $I \times ~\cdots~ \times I$. The
maximum is

\begin{equation} \label{twosixteen}
S^{\max}_{hs} = N \log_2 V\ .
\end{equation} 
The minimum (which occurs for completely fine-grained histories) is zero.

Another useful quantity is the Lloyd-Pagels (LP)
depth \cite{LP88}, defined as

\begin{eqnarray} \label{twoseventeen}
{\cal D}_{LP} (\{c_\alpha\}) &\equiv& S^{\max}_{hs} - S_{hs} (\{c_\alpha\}) 
\nonumber \\
&=& \sum_\alpha p_\alpha \log_2  p_\alpha - \sum_\alpha p_\alpha \log_2
\left[\tr ({\bf P}_\alpha) /  \tr ({\bf I})\right] \ .
\end{eqnarray} 
This has a number of useful features. It is a direct measure of the
information in a set of histories; it is invariant under dimensional
transformations; and it is invariant under refinement of the fine-grained net of
times.

To illustrate, consider the entropy of a history consisting of a set of
alternatives $\{P_\alpha\}$ at a single moment of time $t$. This is 

\begin{equation} \label{twoeighteen}
S_{hs} (\{P_\alpha\}) = -\sum_\alpha p_\alpha \log_2  p_\alpha + 
\sum_\alpha p_\alpha
\log_2\left[\tr(P_\alpha)\right] + (N-1) \log_2 V\ .
\end{equation} 
This is the entropy that would be obtained from the usual Jaynes construction
(\ref{onethree}) with the addition of the constant
$(N-1) \log_2 V$ representing the
missing information at all the other moments of time.  By contrast, the depth

\begin{equation} \label{twonineteen}
{\cal D}_{LP} (\{P_\alpha\}) = \sum_\alpha  p_\alpha \log_2  p_\alpha - 
\sum_\alpha p_\alpha
\log_2\left[\tr({\bf P}_\alpha) / \tr({\bf I})\right]
\end{equation} 
is the same as the $-S$ that would be calculated from (\ref{onethree}),
without extra terms.
Note that if we use a dimensionally invariant form $S_{hs}$, 
by subtracting a term $\log_2 V^N$ as suggested above,
we would have the simple relationship
\begin{equation}
S_{hs}(\{c_\alpha\}) - \log_2 V^N = - {\cal D}_{lp}(\{c_\alpha\}).
\end{equation}

\subsection{Other Entropies of Histories}

The history space entropy is not the only information measure that can be
associated with histories. In the following we discuss some others and the
relationships between them.

\noindent
{\sl Isham and Linden's Entropies}.

In their seminal paper on entropy in generalized quantum theory
\cite{IL97}, Isham and Linden utilize history space to define a one
parameter family of entropies based on the decoherence functional
$D(\alpha, \alpha')$ for a decoherent set of coarse-grained
histories. Translated into the notation of this paper their definition
reads:
\begin{equation}\label{A} 
I_x(\{c_\alpha\})=-\sum_\alpha p_\alpha \log_2 p_\alpha + x \sum_\alpha
p_\alpha \log_2 \left[\tr ({\bf P}_\alpha)/\tr ({\bf I})\right]\ .
\end{equation}
As they show explicitly, for $x\ge 1$ the entropy $I_x(\{c_\alpha\})$
possesses the important property that it increases under coarse-graining
of the decoherent set. 

As discussed by Isham and Linden, in the case of non-relativistic
quantum mechanics, history 
space is a repeated tensor product of the Hilbert space of the system --- one
factor for each time. The ${\bf P_\alpha}$ are projections on this space
and the trace is defined as usual. However, their arguments can be 
immediately applied to the classical situations we have been discussing.
 (We shall return to the quantum mechanical case in Section V.) 
Indeed, any classical problem can be considered as a generalized quantum
theory in which all sets of alternative histories decohere
automatically: $D(\alpha,\alpha')\equiv p(\alpha)\,
\delta_{\alpha\alpha^\prime}$. The expression
(\ref{A}) thus applies immediately in
the classical case. The history space entropy
$S_{hs}$ we arrived at from the Jaynes construction corresponds to $x=1$,
up to a possible overall renormalization.
Isham and Linden mainly consider  $x=2$, but 
that should not obscure the fact that our history space entropy, 
defined through a Jaynes construction, is a special case
of those that they consider. In Section V we will provide a Jaynes
construction for this entropy in quantum mechanics. 

\vfill\eject
\noindent
{\sl Step-by-step entropy}.

Consider the special case where the set of coarse-grained
histories consists of a sequence of
sets of coarse-grained
alternatives at a series of times $t_1,~\cdots,~t_n$. For generality we
assume that these sets are branch dependent, that is, the sets 
at time $t_k$ may depend on the specific choice of sets at previous times
$t_1\cdots t_{k-1}$. The projections at time $t_k$ then have the form
\begin{equation} \label{twotwenty}
P^k_{\alpha_k} (\alpha_{k-1}, ~\cdots,~\alpha_1) \ .
\end{equation} 
At any stage in the sequence $k=1, ~\cdots, ~n$, one can construct the Jaynes
entropy of the set of alternatives $\{P^k_{\alpha_k}\}$
conditional on a particular
previous history $\alpha_{k-1}, ~\cdots,~\alpha_1$. This is, following
(\ref{twoeighteen}),
\begin{eqnarray} \label{twotwentyone}
S_k\left(\{P^k_{\alpha_k}\} | \alpha_{k-1}, \cdots, \alpha_1\right)  =
&-& \sum_{\alpha_k} p_{\alpha_k|\alpha_{k-1}, \cdots, \alpha_1} \log_2
p_{\alpha_k|
\alpha_{k-1}, \cdots, \alpha_1} \nonumber \\
&+&\sum_{\alpha_k} p_{\alpha_k|\alpha_{k-1}, \cdots, \alpha_1} \log_2
\left\{\tr\left[P^k_{\alpha_k}~(\alpha_{k-1}, \cdots,
\alpha_1)\right]\right\}\ .
\end{eqnarray} 
where $~p_{\alpha_k|\alpha_{k-1}, \cdots, \alpha_1}$
is the conditional  probability
for $\alpha_k$ given the previous history
$\alpha_{k-1}, \cdots, \alpha_1$. In terms of
joint probabilities this is 

\begin{equation} \label{twotwentytwo}
p_{\alpha_k|\alpha_{k-1}, ~\cdots,~\alpha_1} ~=~ {~p_{\alpha_k, \cdots, \alpha_1} \over
p_{\alpha_{k-1}, \cdots, \alpha_1}} \ .
\end{equation} 

Average the conditioned entropies (\ref{twotwentyone}) over past histories
weighted by their probabilities and sum over all the steps from one to $n$ to
obtain the step-by-step entropy $S_{sbs}
\, (\{P^n_{\alpha_n}\}, \cdots, \{P^1_{\alpha_1}\})$:

\begin{eqnarray} \label{twotwentythree}
S_{sbs} \left(\{P^n_{\alpha_n}\}, \cdots,
\{P^1_{\alpha_1}\}\right) 
 &=& 
\sum^n_{k=1} 
\ \sum_{\alpha_{k-1}\cdots\alpha_1}p_{\alpha_{k-1}\cdots\alpha_1} 
  S_k\left(\{P^k_{\alpha_k}\} ~|~ \alpha_{k-1}~\cdots~\alpha_1\right) \ .
\end{eqnarray} 
A little algebra using (\ref{twotwentytwo}) and
(\ref{twotwentythree}) is enough to show that, for the case of
sets of alternatives at a series of times,
the step-by-step entropy and the
history space entropy are related by
\begin{equation} \label{twotwentyfour}
S_{hs}(\{c_\alpha\}) = S_{sbs} \left(\{P^n_{\alpha_n}\}, \cdots,
\{P^1_{\alpha_1}\}\right)+ (N-n) ~\log_2 V\ .
\end{equation} 
That is, they are identical except for a constant factor that is the missing
information at the times not specified. Evidently, $S_{sbs} < S_{hs}$ for the
coarse-grained sets for which they are both defined.

\subsection{Dynamically Constrained History Space Entropy}

In constructing of the history space entropy $S_{hs}$, 
the entropy functional (\ref{twoeleven}) is maximized over all
probability functions ${\bf \widetilde{W}} 
({\bf x})$ irrespective of whether they
conform to the same basic dynamical law. $S_{hs}$ 
is thus the missing information in
histories assuming we are also missing any information about the dynamics.

A dynamical law could be enforced by maximizing
$\cs(\widetilde{\bf W})$ only over the
$\widetilde{\bf W}$ that conform to it.
For example, by maximizing over 
the form (\ref{twoseven}), keeping $p_\eta(x | y)$ fixed,
we enforce a particular
Markovian dynamical law. The resulting entropy
$S_{dc}(\{c_\alpha\})$ we call
the {\it dynamically constrained history space entropy}.
The maximum in (\ref{twotwelve}) is carried out
only over the initial distribution $\tilde{\rho}(x)$ with ${\bf \widetilde W}
(x)$ determined by enforcing the subsequent dynamics explicitly.
Evidently, since this is a constrained maximum,

\begin{equation} \label{twotwentysix}
S_{dc}(\{c_\alpha\}) \leq S_{hs}(\{c_\alpha\})\ .
\end{equation} 

The entropy $S_{dc}(\{c_\alpha\})$ is connected to another entropy of histories 
obtained by
applying the Jaynes method used in (\ref{onethree}) to the initial
${\tilde\rho}(x)$,  but constraining the maximum
not simply by the requirement that probabilities at one time are reproduced,
but probabilities of a whole set of histories.
We call this the {\it initial condition entropy} $S_{ic} (\{c_\alpha\})$.

We can illustrate the
construction of $S_{ic}$
in the case of  Markovian evolution and a set of histories that is
a sequence of sets of alternatives $\{P^k_{\alpha_k}\}$ at a series of times
$t_k,~k=1,~\cdots, ~n$. The probabilities of these histories are given by
(\ref{twoten}) which we may conveniently write as

\begin{equation} \label{twotwentyseven}
p_{\alpha_n ~\cdots ~\alpha_1} = \tr \left(C_{\alpha_n \cdots\alpha_1} 
\rho \right)
\end{equation} 
that is, the sum or integral of $\rho(x_0)$ with the functions $C_{\alpha_n ~\cdots~\alpha_1} (x_0)$ defined 
by (\ref{twoten}). We can
now carry out the Jaynes construction

\begin{equation} \label{twotwentyeight}
S_{ic} (\{c_\alpha\}) 
= \max_{\tilde{\rho}} ~\cs (\tilde{\rho}) \mid_{
\tr \left[C_{\alpha_{n}~\cdots~\alpha_1}\tilde{\rho}\right]
= \tr\, \left[C_{\alpha_{n}~\cdots~\alpha_1}\rho\right]}
\end{equation} 
for all the histories.  The density function which
realizes this maximum has the form

\begin{equation} \label{twotwentynine}
\tilde{\rho}(x) ~=~ \exp \left[-\sum_{\alpha_{n}~\cdots~\alpha_1}
~\lambda^{\alpha_{n}~\cdots~\alpha_1}~C_{\alpha_{n}~\cdots~\alpha_1}(x)\right]
\end{equation} 
where the Lagrange multipliers $\lambda^{\alpha_n \cdots \alpha_1}$
are determined by the conditions

\begin{equation} \label{twothirty}
\tr \left(C_{\alpha_n ~\cdots~\alpha_1} \tilde{\rho}\right) = 
\tr \left(C_{\alpha_n
\cdots \alpha_1} \rho \right)\ .
\end{equation} 

The $C_{\alpha_{n}~\cdots~\alpha_1}(x)$ are not projections,
and there seems no easy
way to evaluate (\ref{twotwentynine}) and (\ref{twothirty}) explicitly in
general. However, $S_{dc}$ and $S_{hs}$
supply upper bounds on $S_{ic}$ as we shall now show.

Write out the entropy functional (\ref{twoeleven}) for 
${\bf \widetilde W}$ of the form
(\ref{twoseven}) to find after a little algebra

\begin{equation} \label{twothirtyone}
\cs(\tilde{\rho}) ~=~ \cs({\bf \widetilde{W}}) ~-~ \int dx_0 ~s(x_0) 
~\tilde{\rho}(x_0)
\end{equation} 
where $\cs({\bf \widetilde{W}})$ is (\ref{twoeleven}) for 
${\bf W}$ of the
form (\ref{twoseven}), and 

\begin{equation} \label{twothirtytwo}
\cs(\tilde{\rho}) ~=~ - \tr (\tilde{\rho} ~\log_2 ~\tilde{\rho}) \ .
\end{equation} 
The entropy $s(x_0)$ is defined by 

\begin{equation} \label{twothirtythree}
s(x_0) ~=~ -\int dx_n ~\cdots~dx_1
  ~p(x_n ~\cdots x_0) ~\log_2 ~p(x_n ~\cdots x_0)\ ,
\end{equation} 
where

\begin{equation} \label{twothirtyfour}
p(x_n, \cdots, x_0) = p(x_nt_n ~|~x_{n-1}t_{n-1}) ~\cdots~p(x_1 t_1
~|~x_0 t_0)\ .
\end{equation} 
The function $s(x_0)$ is always positive. Thus

\begin{equation} \label{twothirtyfive}
\cs(\tilde{\rho}) ~\leq ~\cs({\bf \widetilde{W}})
\end{equation} 
for ${\bf \tilde{W}}$ of the form (\ref{twoseven}).  [Note that
the Markovian form of the dynamics in (\ref{twothirtyfour}) is not important;
any probability function $p(x_n,\ldots,x_0)$ satisfies this
inequality.]  On maximization over $\tilde{\rho}$ we have the inequalities

\begin{equation} \label{twothirtysix}
S_{ic} ~\leq~ S_{dc} ~\leq~ S_{hs}\ .
\end{equation} 
In particular, if on fine-graining $S_{hs}$ is driven to a low value, then
$S_{dc}$ and $S_{ic}$ will be as well. 
We shall use this in what follows.

\section{Behavior of History Space Entropy Under Fine-Graining}

Entropies decrease under fine-graining and increase under coarse-graining.
That immediately follows from the Jaynes construction as the discussion
leading to (\ref{twofourteen}) shows. Usually this well-known behavior is
considered for variations in levels of coarse-graining at a given moment of
time.  However, histories can also be fine-grained {\it in} time. 
For example, if
a set of histories is specified by one set of alternatives at a series of
times, and another set of histories
by the same alternatives at {\it more} times, then the second set
is a fine-graining of the first.

In this section we examine explicitly the behavior of history space entropy
under fine-graining in three one-dimensional models with simple stochastic
evolutionary laws. They are: a discrete random walk, continuous diffusion, and
Brownian motion. The random walk is the simplest model; diffusion
illustrates the modifications necessary for the continuum; and Brownian
motion is a simple example of a non-Markovian process.  In all cases we
consider a finest-grained net of $N$ equally-spaced times so that
fine-grained histories are specified by $N$ positions $(x_1,\cdots, x_N)$.
We consider coarse-grainings in which these positions are grouped into
equal intervals of size $\Delta x$ at a series of times
spaced by equal intervals $\Delta t$.  
We then study history space entropy for these
coarse-grainings as a function of $\Delta x$ and $\Delta t$.

\subsection{Random Walk}

We take an initial condition where all
histories begin at the initial point $x_0 = 0$ and assume that
at each timestep
the particle has an equal chance of moving right ($x\rightarrow x+1$) or
left
($x \rightarrow x-1$) on a discrete spatial lattice.  There are then  
$2^N$ fine-grained histories 
with equal probability $1/2^N$ and all other histories have probability $0$.  
We assume that
the lattice has a large finite size $V$ with periodic boundary conditions
relating its ends. The history space entropy is given by (\ref{twofifteen})
where $\tr({\bf P}_\alpha)$ is the number of fine-grained histories in a
coarse-grained history $c_\alpha$.  For all histories with the
coarse-graining described above this is:
\begin{equation}
\tr({\bf P}_\alpha) = (\Delta x)^{N/\Delta t} V^{N(1-1/\Delta t)}\ .
\label{threeone}
\end{equation}

Simple as this is, it is clear that as the number of fine-grained histories
increases rapidly with the number of times $n=N/\Delta t$, and calculating
entropies by summing over all the fine-grained histories in each coarse-grained
history rapidly becomes impractical.
Instead, we use a Monte Carlo approach:  we generate a
large sample of fine-grained histories, bin them together into
coarse-grained classes, and calculate the entropies from the resulting
probability estimates.  This technique works in the continuous
case as well.

In Figure 1 we plot the history space
entropy $S_{hs}$ of the random walk model as a
function of the $\Delta x$ and $\Delta t$. We clearly see
that the entropy rises steeply when the coarse-graining is increased by
increasing $\Delta t$ and more moderately as $\Delta x$ is increased.
Increasing $\Delta x$ to $V$ at a fixed time gives the maximal 
coarse-graining where the only
alternatives are $(I, 0)$.
Therefore, increasing $\Delta x$ to $V$
for any fixed value of $\Delta t$ will give the maximum possible entropy,
associated with the alternative ${\bf P}={\bf I}$. That is, from
(\ref{threeone}) (the $-\sum p \log_2\, p$ term vanishes),
\begin{equation}
S^{\rm max}_{hs} = N\, \log_2\, V\ .
\label{threetwo}
\end{equation}

\centerline{\epsfysize=5.50in \epsfbox{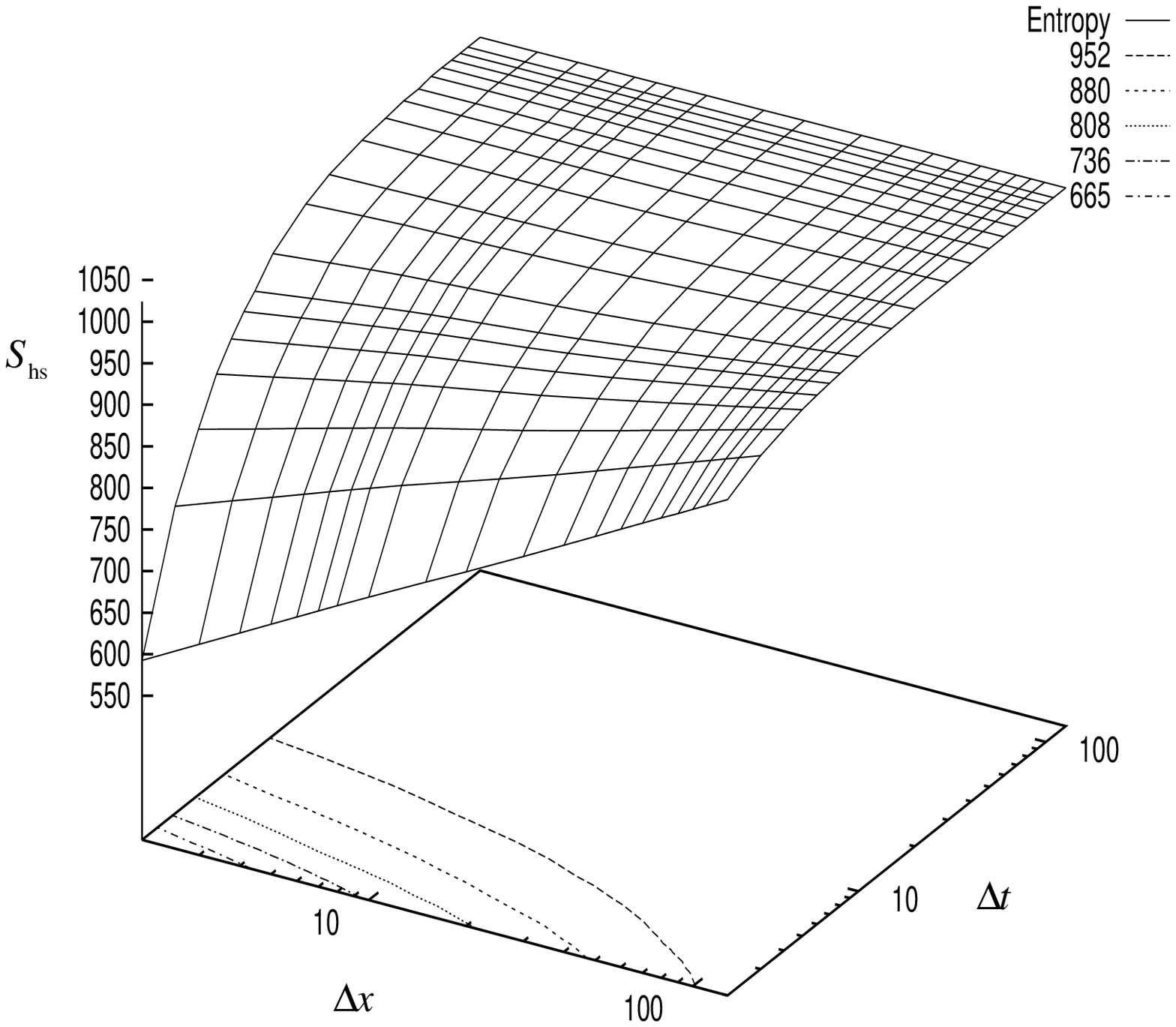}}

\begin{quote}{{\bf Figure 1:} \sl History space entropy, $S_{hs}$, for the
discrete random walk as a function of coarse-graining scales $\Delta x$ and
$\Delta t$. In this system, a particle begins at $x=0$ on a 1D lattice of
256 points and moves left or right by 1 position with equal probability at
each of $N=128$ times. The entropy is measured in bits of missing
information. These results were produced by a Monte Carlo simulation with
100,000 random trajectories; because of the rapid rise in $S_{hs}$ with
coarse-graining, the $\Delta x$ and $\Delta t$ axes are plotted on a
logarithmic scale.}
\end{quote}

Increasing $\Delta t$ for fixed $\Delta x$ gives a closely related limit.
When $\Delta t$ is at its maximum value of $N$, $S_{hs}$ is the single time
entropy plus $(N-1)\, \log_2\, V$ [{\it cf.} (\ref{twotwentyfour})]. The
single time entropy ranges from $N\, \log_2\, 2$ for $\Delta x=1$ to $N\,
\log_2\, V$ for $\Delta x=V$. Thus, for large $N$ we expect $S_{hs}$ to be
essentially $S^{\rm max}_{hs}$ and that behavior is also illustrated in
Figure 1.

The maximum value of $S_{hs}$ for the
particular model simulated is $N\log_2 V$, which in this case is 1024 bits.
This is reflected on the plot. At the other extreme, the minimum entropy
occurs for $\Delta t= \Delta x=1$, and is $S_{hs} =128$ bits. The
finest graining included in Figure 1 is $\Delta t= \Delta x = 2$, and we
see that $S_{hs}$ has already risen steeply at that point.

\subsection{Continuous Diffusion}

A Markovian diffusion process illustrates the case when $M$ is a continuous
space.  Take the transition probability to be 
\begin{equation}
p(x_2,t_2|x_1,t_1) = {1\over\sqrt{\pi D\Delta t}}
  \exp[ - (x_2-x_1)^2/D\Delta t ]\ ,
\label{continuous_walk}
\end{equation}
where $D$ is a diffusion constant and $\Delta t = t_2 - t_1$.
Assume a finite range size $V$, divided into cells of size
$\Delta x$, and a total duration for the histories of $t_f =
N\Delta t$.  Choose $V \gg \sqrt{Dt_f}$, so that we  needn't worry
about boundaries. We again assume an initial condition where the particle
is initially at $x_0$.

Label the intervals of the spatial coarse-graining by an integer $i$, a
point lying in the $i^{\rm th}$ cell if $i\Delta x \le x < (i+1)\Delta x$.
The probability of a particle initially at $x_0$ passing
through a sequence of $n$ cells $i_1, \ldots, i_n$ at times
$t_j = j\Delta t$ is 
\begin{eqnarray}
p(i_1, \ldots, i_n) & =& \int_{i_1\Delta x}^{(i_1+1)\Delta x} dx_1
  \cdots \int_{i_n\Delta x}^{(i_n+1)\Delta x} dx_n
  \prod_{j=1}^n p(x_j,j\Delta t|x_{j-1},(j-1)\Delta t) \nonumber \\
& =& {1\over(\pi D\Delta t)^{n/2}} \int_{i_1\Delta x}^{(i_1+1)\Delta x}
dx_1
  \cdots \int_{i_n\Delta x}^{(i_n+1)\Delta x} dx_n
  \exp\left[ - \sum_{j=1}^n {(x_j-x_{j-1})^2\over D\Delta t} \right].
\label{threefour}
\end{eqnarray}
The history space
entropy for the continuous case has exactly the same
form as (\ref{twofifteen}) the discrete case, but it is convenient to
make use
of a dimensionally invariant form of the entropy, by subtracting 
a dimensional factor $\log_2 V^N$.  Thus, the $\log_2\tr({\bf P}_\alpha)$
term in (\ref{twofifteen})
becomes $\log_2[\tr({\bf P}_\alpha)/\tr({\bf I})]$. Rather than
being an integer, as the discrete case, it is
a continuous  
measure of the coarse-graining of each history.  For the coarse-graining
described above, with intervals of size $\Delta x$ and $n=t_f/\Delta t$ 
times, we get
\begin{equation}
\log_2 \left[\tr({\bf P}_\alpha)/\tr({\bf I})\right] = n\log_2(\Delta x/V)\ .
\label{threefive}
\end{equation}
$\Delta x/V < 1$, so $\log_2[\tr({\bf P}_\alpha)/\tr({\bf I})] < 0$
for all but maximally
coarse-grained histories.  

There are $(V/\Delta x)^n$ coarse-grained histories.  The $p\log_2 p$
part of the entropy in (\ref{twofifteen})
is maximized in the case when all the histories
have equal probabilities.  In this case,
\begin{equation}
\max \sum_\alpha \left(- p_\alpha \log_2 p_\alpha\right)  = 
n\log_2(V/\Delta x)\ ,
\label{threesix}
\end{equation}
Taking account of (\ref{threefive}) we see that $0$ is the maximum of
$S_{hs}$ so that it is strictly non-positive.
This is different from usual definitions of entropy,
which are logarithms of large numbers and hence always positive.
However, what is important is the change
in $S_{hs}$ under coarse-graining or refinement, not its absolute value.

We can gain some insight by looking at the limiting behavior of $S_{hs}$
for different levels of coarse-graining.  Consider first the
coarse-grained limit where $\Delta x \rightarrow V$.  As $\Delta x$
becomes large compared to $\sqrt{Dt_f}$, it becomes highly improbable
that the particle will ever diffuse outside of a single cell $i$.
Thus, in this limit, one history dominates with a probability $p \approx 1$
while the others are suppressed, $p \approx 0$, and the $-\sum p\ \log_2 
\ p $ part of the entropy vanishes. 
At the same time, the term
$\log_2 [\tr({\bf P}_\alpha)/\tr({\bf I})] = n\log_2(\Delta x/V)$ approaches 0
as well, so this maximal coarse-graining in $x$ leads to
$S_{hs} \rightarrow 0$;  $S_{hs}$ is maximized by maximal
coarse-graining in $x$.

Let us go now to the opposite limit, where $\Delta x \ll V$.  We can
now label the interval $i_j$ by the value $x_j$ centered in that interval.
The probability to go from $x_{j-1}$ to $x_j$ is
\begin{equation}
p(x_j|x_{j-1}) = {\Delta x\over\sqrt{\pi D\Delta t}}
  \exp\left[ - {(x_j-x_{j-1})^2\over{D\Delta t}} \right].
\label{threeseven}
\end{equation}
The $p\log_2 p$ term for a single history is then
\begin{eqnarray}
- p(x_1,\ldots,x_n) \log_2 p(x_1,\ldots,x_n) =
  - {n\over2}\log_2\left({\Delta x^2\over{\pi D\Delta t}}\right)
  p(x_1,\ldots,x_n) \nonumber\\
+ (\log_2 e) \left( \sum_j {(x_j-x_{j-1})^2\over{D\Delta t}} \right)
   \left({\Delta x^2\over{\pi D\Delta t}}\right)^n
   \exp\left[ - \sum_j {(x_j-x_{j-1})^2\over{D\Delta t}} \right]\ .
\label{threeeight}
\end{eqnarray}
Summing over all histories is the same as summing the above expression
over all the $x_j$.  These sums can be approximated by integrals,
which are readily evaluated to yield
\begin{equation}
\sum_\alpha \left(- p_\alpha \log_2 p_\alpha\right) \approx
  n \log_2\left({\sqrt{\pi Dt_f}\over\Delta x}\right)
  - {n\over2}(\log_2 n - 1)\ .
\label{threenine}
\end{equation} 
Adding the expression for
$\log_2 \tr({\bf P}_\alpha)/\tr({\bf I})$ from (\ref{threefive})
gives for the entropy
\begin{equation}
S_{hs}  \approx
  n \log_2\left({\sqrt{\pi Dt_f}\over V}\right)
  - {n\over2}(\log_2 n - 1) < 0\ ,
\label{threeten}
\end{equation}
{\it i.e}., $S_{hs}$ approaches a constant negative value in the limit of 
small
$\Delta x$ for a fixed $\Delta t$.

Suppose now that we hold $\Delta x$ fixed and vary the coarse-graining
in $t$.  If we go to the maximum coarse-graining $\Delta t = t_f$,
we return to the case of alternatives at a single time.  If the probability
of the particle being in the interval $i$ is $p_i$, the entropy is just
\begin{equation}
S_{hs} = - \sum_i p_i \log_2 p_i + \log_2(\Delta x/V)\ ,
\label{threeeleven}
\end{equation}
differing from the usual single-time entropy only by a constant.

\centerline{\epsfysize=5.50in \epsfbox{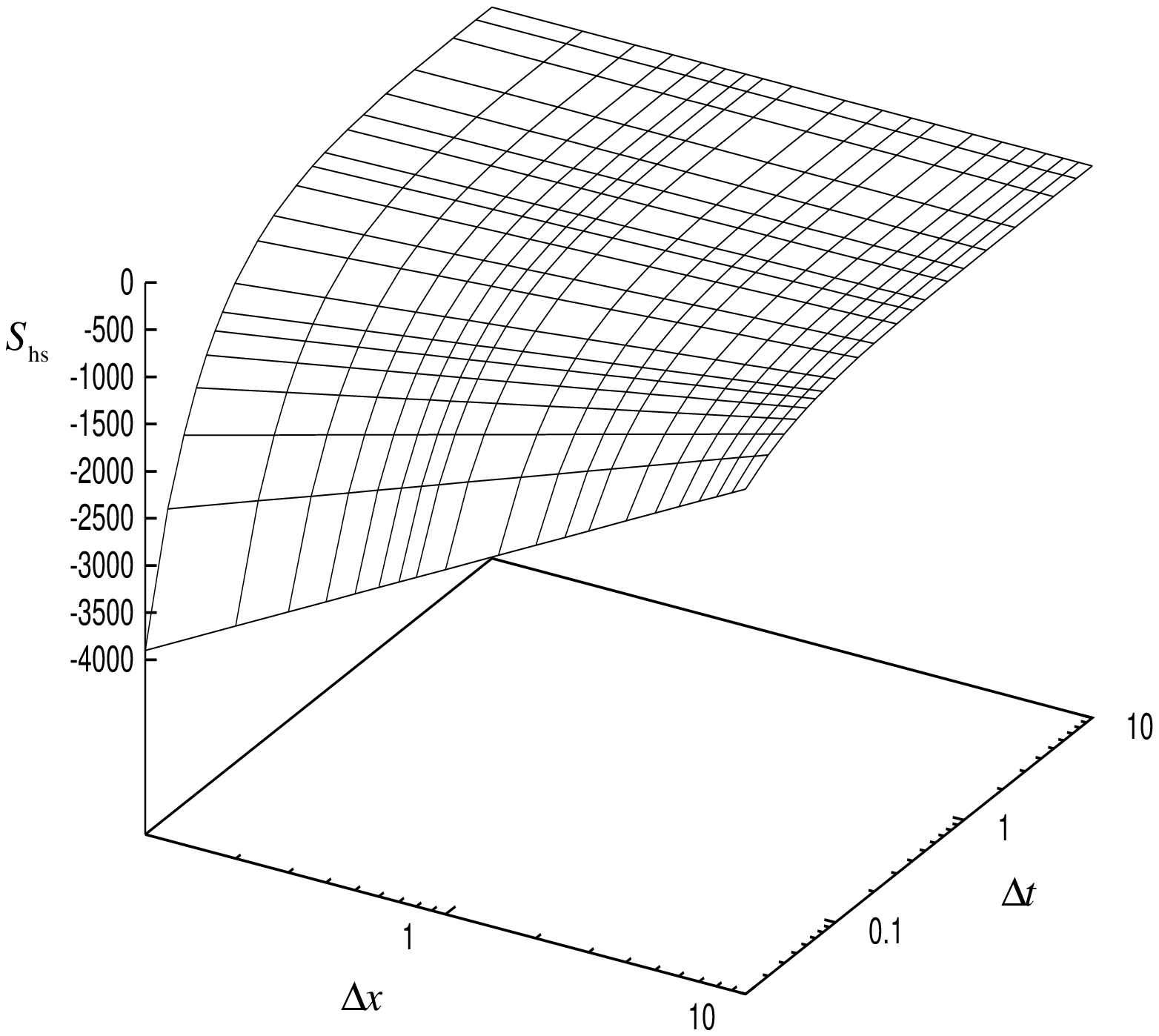}}

\begin{quote} {{\bf Figure 2:} \sl History space entropy $S_{hs}$ for
continuous diffusion as a function of coarse-graining scales $\Delta x$ and
$\Delta t$, in bits.
All particles begin at $x=0$ on a 1D manifold of length $V=20$,
and spread with diffusion constant $D=1$ through a finest-grained net of
$N=1024$ times with minimal timestep $\eta=0.01$. The finest-grained cell
size is $\Delta x=0.1$.
We have subtracted off the the maximum entropy $N \log_2 V$ to render our
results invariant under dimensional rescaling and refinements in time; the
maximum entropy is thus $0$, and $S_{hs}$ is not bounded below. These
results were produced by a Monte Carlo simulation with 10,000 random
trajectories.}
\end{quote}

If instead we refine the description in time the result is quite
different.  As the timestep $\Delta t$ becomes small compared to
$\Delta x^2/D$, the probability of a particle moving from one interval
to another in a timestep becomes small as well.  Beyond that point,
refining the description of the system in time does not increase the
actual number of alternative histories with non-zero probabilities.  Thus,
\begin{equation}
- \sum p \log_2 p \rightarrow {\rm const.}
\label{threetwelve}
\end{equation}
The $\log_2\, [\tr({\bf P}_\alpha)/\tr({\bf I})] = n \log_2(\Delta x/V)$
term, however, does change as we increase
$n$.  Because this term is negative, as we increase
the fine-graining in $t$, the history space entropy $S_{hs}$ decreases
without limit.

In both $x$ and $t$, the entropy is diminished by making the description
more fine-grained.  Thus, we expect the same behavior as in the
simple random walk:  the entropy $S_{hs}$ (and thus, all other measures
of entropy for histories that we have considered) will be minimized
by the most fine-grained description.
We performed a numerical calculation to generate the entropy
plot in Figure 2.  Note that the qualitative behavior is exactly
the same as in Figure 1.  

\subsection{Brownian motion}

In the previous examples, we assumed an explicitly Markovian
time-evolution.  If we relax that assumption and
suppose that the probability of a
history $p(x_1,\ldots,x_n)$ does not have the form
$p(x_n|x_{n-1}) \cdots p(x_2|x_1) p(x_1)$ are our conclusions affected?

As a simple example of a non-Markovian process, consider a
particle undergoing Brownian motion.  In
addition to inertia and dissipation, the particle is subjected to
a stochastic force.  We can write a stochastic differential
equation for its motion in It\^o form:
\begin{eqnarray}
dx = && (p/m) dt\ , \nonumber\\
dp = && - 2\Gamma p dt + a d\xi\ ,
\end{eqnarray}
where $d\xi$ is a stochastic differential variable with zero mean
and variance $dt$,
\begin{equation}
{\rm M}(d\xi) = 0,\ \ {\rm M}(d\xi^2) = dt\ .
\end{equation}
This stochastic equation corresponds to a Fokker-Planck equation
for probability densities $\rho(x,p,t)$ in phase space \cite{HE80}:
\begin{equation}
{d\rho\over dt}(x,p) = - \left(\frac{p}{m}\right) 
{\partial\rho\over\partial x}(x,p)
  + 2 \Gamma {\partial\over\partial p} p \rho(x,p)
  + {a^2\over2} {\partial^2\rho\over\partial p^2}(x,p)\ .
\label{threefourteen}
\end{equation}

We can enumerate a set of coarse-grained histories for Brownian
motion just as we did for the continuous random walk, dividing up
the range $V$ into cells of size $\Delta x$ and dividing the
total time of the histories $t_f$ into $n$ steps of $\Delta t$ each.
An individual coarse-grained history consists of
all fine-grained histories which pass through a
given set of intervals $i_1,\ldots,i_n$ at times $t_j = j\Delta t$.

Histories of $x(t)$ are not Markovian because of the existence
of the inertia term $-(p/m)\partial\rho/\partial x$ in (\ref{threefourteen}).  
However, looked at over relatively long
times $\Delta t \gg 1/\Gamma$ the inertia becomes unimportant, as
dissipation dominates.  On these
long timescales, the system is well approximated by the
continuous diffusion model (\ref{continuous_walk}) with
$D=a^2/8\Gamma^2 m^2$.
On very short timescales, by contrast, inertia dominates.  The
particle drifts at a near-constant velocity, only slightly deflected
by dissipation and noise.

We see that the same arguments we used in the case of
continuous diffusion apply to this case with little
modification.  Fine-graining in $t$ reduces the entropy without
limit.  Fine-graining in $x$ is a little less clear, but a
similar argument can be made.  In the limit of fine-grained $x$,
we can approximate the probability of a history as
\begin{equation}
p(x_1,\ldots,x_n) = (\Delta x/Q_0)^n f(x_1,\ldots,x_n)\ ,
\end{equation}
where $f(x_1,\ldots,x_n)$ is dimensionless, and $Q_0$ is a constant
with units of length which depends on $\Delta t$ but not $\Delta x$.
We can replace the sum over all histories in (\ref{twofifteen})
with $n$ integrals over the $x_j$, and get
\begin{eqnarray}
S_{hs} = &&
 - {1\over Q_0^n}\int dx_1\cdots dx_n\ f(x_1,\ldots,x_n) \log_2 f(x_1,\ldots,x_n)
\nonumber\\
&& - n \log_2(\Delta x/Q_0) + n \log_2(\Delta x/V) \nonumber\\
\equiv && S_0 + n \log_2(Q_0/V)\ ,
\end{eqnarray}
where $S_0$ has no $\Delta x$ dependence.
Since the $\Delta x$ dependence has dropped out completely,
we see that in this case as well the entropy approaches a constant as
we fine-grain in $x$.

In Figure 3 we show the numerical results for the entropy of coarse-grained
histories of the Brownian motion model as a function of coarse-graining
in $x$ and $t$.  This graph clearly shows essentially the same behavior
of $S_{hs}$ with coarse-graining as in Figures 1 and 2.

\vskip .13 in
\centerline{\epsfysize=5.50in \epsfbox{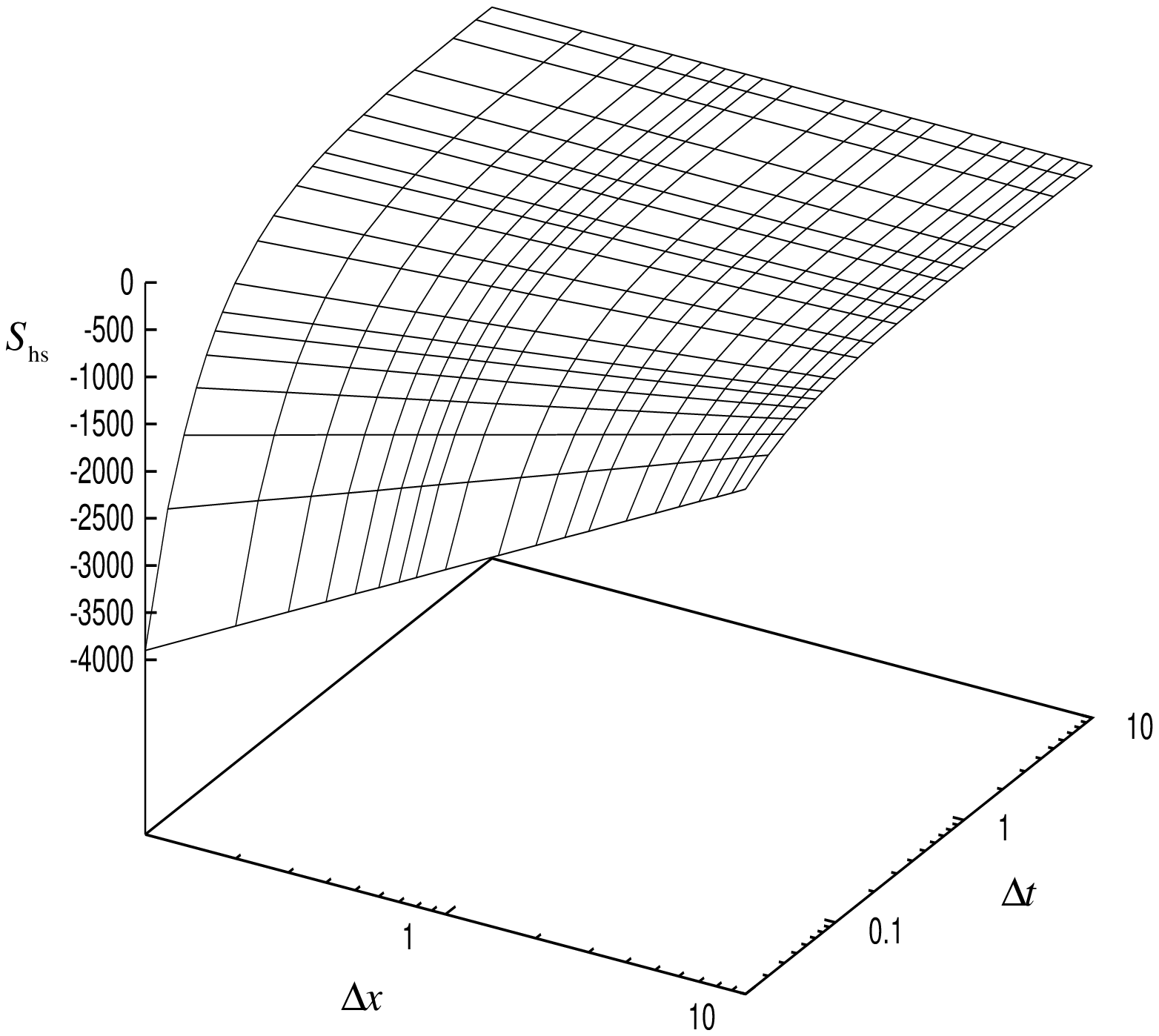}}

\begin{quote} {{\bf Figure 3:} \sl History space entropy, $S_{hs}$, for
Brownian motion as a function of coarse-graining scales $\Delta x$ and
$\Delta t$, in bits. All particles begin with $(x,p)=(0,0)$
on a 1D manifold of length $V=20$,
with dissipation $2\Gamma=1$, noise strength $a=1$, and mass $m=1$.
The finest-grained net of times and minimum cell size are as in Figure 2,
and the same conventions are used here in displaying $S_{hs}$.
These results were produced by a Monte Carlo simulation with 10,000 random
trajectories.}
\end{quote}

\section{The Second Law for Histories}

\subsection{The Increase of Entropies}

The familiar second law of thermodynamics concerns the behavior of the
entropy of a fixed set of coarse-grained alternatives at a moment of time
as this time is varied.  We shall call such entropies ``single-time
entropies''. 

If the value of a single-time entropy at some particular time $t_0$ is all
that is known about a system, and if that value is much lower
than the maximum (equilibrium) value, then that entropy will
subsequently tend to increase
for most dynamical laws of interest.  If the
dynamical law is time symmetric about $t_0$, then the approach to
equilibrium will also be symmetric about $t_0$.  However, it is not just
this statistical tendency to approach equilibrium that is usually
meant by the second law of thermodynamics. Rather, it is the general
increase in entropy of suitable coarse-grained descriptions of the
universe since the big bang. In particular, what is meant is that, for the most
part,  certain entropies of presently isolated systems
are increasing in the same direction of time.
The time-asymmetric increase of these entropies of the universe
arises from a cosmological initial condition at which those entropies
were low.
As Boltzmann put it ``The second law of thermodynamics can be proved from
the mechanical theory if one assumes that the present state of the universe
\dots\  started to evolve from an improbable state'' \cite{Bol97}..

The entropies that are most useful in chemistry and physics are
associated with quasiclassical coarse-grainings which fix the 
values of averages over suitable volumes of densities of approximately
conserved quantities such as energy, momentum, and abundances of
chemical and nuclear species. Their utility arises from the approximate
conservation. The small volumes over which the averages are taken 
reach local equilibrium on short time scales, leaving the approach
to equilibrium between volumes to be described by phenomenological
equations such as the Navier-Stokes equation over longer time scales. 
The single-time entropy of these coarse-grainings is low in the 
early universe leading to a general tendency to increase. 

Statements of the second law often refer to the increase of ``the'' entropy
as though there were only one possible
coarse-grained description for which it
holds. What is meant by 
``the'' entropy is usually the single-time entropy of the
alternatives defining the quasiclassical realm of everyday experience
described above.
However, we should expect the general increase of the entropy of 
{\it any} set of
coarse-grained alternatives which is low in the initial moments of the
universe.  To give just one example, the single-time entropy of a set of
quasiclassical alternatives $\{P_\alpha\}$ increases with time when
conditioned on various other quasiclassical alternatives $\{P_\beta\}$.
Indeed, such entropies
\begin{equation}
S\left(\left\{P_\alpha\right\} ,  t\ |\ \beta, t^\prime\ \right)
\label{fourone}
\end{equation}
are the ones of practical interest.  The entropy of a gas inside a
piston is the entropy of alternatives referring to the gas {\it given} the
configuration of the piston. There are thus a variety of coarse-grainings
and conditions for which the missing information increases with time.

\subsection{The Increase in History Entropies}

Sets of alternative, coarse-grained histories provide more general
coarse-grained descriptions of the universe than sets of coarse-grained
alternatives at merely one time.  The
corresponding entropies of histories should also increase with time if they
are low at the time of the system's initial condition. For example,
consider a set of histories consisting of a series of alternatives
$\{P^n_{\alpha_n}\}, \cdots,\{P^1_{\alpha_1}\}$ at a sequence of time
$t_1, \cdots, t_n$ giving  a histories entropy
\begin{equation}
S_{hs}\left(\{P^n_{\alpha_n}\}, t_n, \cdots,
\{P^1_{\alpha_1}\}, t_1\right)\ .
\label{fourtwo}
\end{equation}
If $S_{hs}$ is initially low, and 
these times are all translated forward by an amount $T$, we would expect
\begin{equation}
S_{hs}\left(\{P^n_{\alpha_n}\}, t_n + T; \cdots ;
\{P^1_{\alpha_1}\}, t_1 + T\right)
\label{fourthree}
\end{equation}
to increase with $T$.

A proof of the second law even for single entropies exists only for highly
idealized situations.\footnote{See, {\it e.g.} \cite{Spo91}}
That is partly because entropy does not monotonically increase but
fluctuates about an increasing trend.  We can therefore hardly expect a
mathematical proof of the increase of (\ref{fourthree}) with $T$. However,
the connection of $S_{hs}$ with the step-by-step entropy supports this in
the following way.

Consider histories consisting of alternatives at just two times $t_1$ and
$t_2$. Then from (\ref{twotwentythree}) and (\ref{twotwentyfour})
\begin{eqnarray}
S_{hs}\left(\{P^2_{\alpha_2}\}, t_2;
\{P^1_{\alpha_1}\}, t_1\right)&=& \sum_{\alpha_1}
p(\alpha_1)\ S\ \left(\{P^2_{\alpha_2}\}, t_2 |\alpha_1,
t_1\right)\nonumber \\
& +& S\ \left(\{P^1_{\alpha_1}\}, t_1\right) + {\rm const}\ .
\label{fourfour}
\end{eqnarray}
where the constant is independent of $t_1, t_2$ and the alternatives.  As
$t_1$ increases, the second term in (\ref{fourfour}) increases.  That is
just the usual second law. The first term can also be expected to increase
as both $t_1$ and $t_2$ move away from a low entropy initial condition,
provided $P^1_{\alpha_1}$ is sufficiently coarse-grained that the initial
condition plays an important role in determining future probabilities.

The sequence of times necessary to specify a set of histories presents a
variety of possibilities for investigating the change in entropy. We have
already discussed a uniform translation of all the times. However, we could
also discuss increasing the separation between the times.  For example, in
the two time case of (\ref{fourfour}), $S_{hs}$ increases as $t_1$ is
fixed and $t_2-t_1$ increases. Indeed, that is just a special case of the
usual second law [{\it cf.}~(\ref{fourone})].

\subsection{The Urn Model}

An exactly soluble model which nicely illustrates the increase in history
space entropy is the urn model of P.~and T.~Ehrenfest \cite{Ehr07}. The
model concerns $2R$ numbered balls, each of which is in one of two urns, $A$
or $B$. The system evolves through $N$ discrete time steps.  At each time a
number from $1$ to $2R$ is chosen and that ball is moved from its present urn
to the other.  Fine-grained histories are specified by giving the urn
containing each ball at each of the $N$ times. A simple kind of
coarse-grained history specifies the number of balls in one urn, say $A$,
at one time $t$.  The kind of multi-time, coarse-grained histories we shall
study are specified by giving the number of balls in $A$, $(n_1, \cdots,
n_n)$ at a sequence of the $N$ times $t_1, \cdots, t_n$.

The probabilities relevant for constructing the entropies can be worked out
\cite{Ehr07,Kac59}. The probability of a transition from one time to the
next is:
\begin{equation}
p(n_{j+1},t_{j+1}|n_j,t_j) = { 2R-n_j \over 2R }\delta_{n_{j+1}, n_j+1}
  + { n_j \over 2R } \delta_{n_{j+1},n_j-1}\ .
\label{urn_trans}
\end{equation}
Given that the number of balls in urn $A$ is $n_0$ at time $t_0$, the
probability that $A$ will contain $n_j$ balls at time $t_j$ is:
\begin{equation}
p(n_j,t_j|n_0,t_0) = (-1)^j 2^{-2R} \sum_{l=-R}^R (l/R)^j
  C_{n_j}^l C_{R+l}^{R-n_0}\ ,
\label{urn_prob}
\end{equation}
where the coefficients $C_k^l$ are defined by the identity
\begin{equation}
(1-z)^{R-l}(1+z)^{R+l} \equiv \sum_{k=0}^{2R} C_k^l z^k\ .
\end{equation}
All the rest of the probabilities we shall need are easily constructed from
(\ref{urn_trans}) and (\ref{urn_prob}). 

Consider, by way of example,   the history space entropy for the set of
histories specified by giving the number of balls in $A$ at two times $t_j$
and $t_{j+m}$ assuming an initial condition in which $n_0$ balls are in $A$
at $t_0$. We call these ``two time histories'' for short. From
(\ref{twofifteen}) this is
\begin{eqnarray}
S_{hs}\left(\left\{n_{j+m}, n_j\right\}\right) &=& - \sum_{n_{j+m}, n_j}
p\left(n_{j+m}, n_j|n_0\right) \ \log_2\ p\left(n_{j+m},
n_j|n_0\right)\nonumber \\
&+& \sum_{n_{j+m}, n_j} p\left(n_{j+m}, n_j|n_0\right)\ \log_2
\ \left(\matrix{2R\cr
                n_{j+m}\cr}\right)
\ \left(\matrix{2R\cr
                n_j\cr}\right)\nonumber\\
&+& (N-2)\ \log_2 \left(2^{2R}\right)\ .
\label{foureight}
\end{eqnarray}
The probability $p(n_{j+m}, n_j|n_0)$ is obtained by multiplying
(\ref{urn_prob}) by a factor of (\ref{foureight}) for each of the $m$ times
between $t_j$ and $t_{j+m}$ and summing over the intermediate values of
$n_k,\ j<k<m$. There are $2^{2R}$ ways of arranging the balls among the urns
at each time so that, and a binomial coefficient gives the number of
arrangements of balls in which $n$ are in urn $A$. Thus,
\begin{equation}
\tr(I) = 2^{2R}\quad ,
  \quad \tr\left(P_n\right) = \left(\matrix{2R\cr
  n\cr}\right)\ .
\label{fournine}
\end{equation}

\centerline{\epsfysize=5.00in \epsfbox{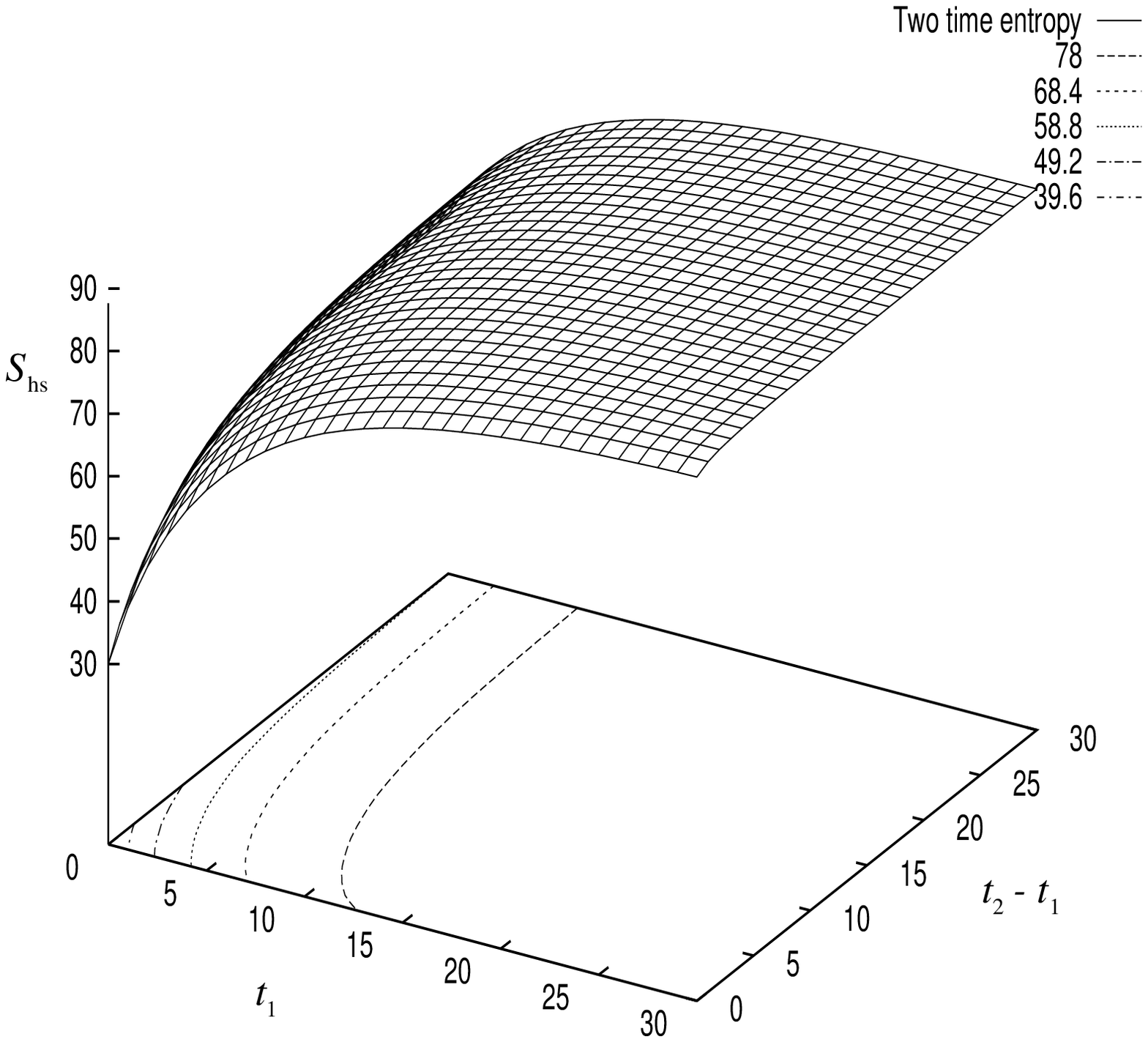}}

\begin{quote} {{\bf Figure 4:} 
\sl History space entropy, $S_{hs}$, for
two time histories of the Ehrenfest urn model as a function of $t_1$
and $m=t_2-t_1$, in bits.  In the case shown there are $2R=30$ numbered balls
distributed between urns $A$ and $B$, with all the balls initially
in urn $A$.  In Figures 4 and 5 we have set the total number of
fine-grained times arbitrarily at $N=3$; a larger, more realistic
number would merely add a constant displacement to $S_{hs}$.}
\end{quote}

It takes of order $2R$ time steps to share information
among the $2R$ balls, and that is the order of characteristic relaxation
time for entropies to increase to their maximum value \cite{Kac59}.  This
is the case for the entropies of
two time histories as $t_1$ and $t_2$ are
increased keeping their difference constant; this was suggested by
(\ref{fourfour}) and shown by Figure 4.
The relation (\ref{fourfour}) shows that the maximum value
(not including the neglected times) is roughly twice the maximum
entropy for single time coarse-grainings of this type.

This relation also indicates that $S_{hs}$ should grow with the
same characteristic relaxation time as $t_2-t_1$ is increased, keeping
$t_1$ fixed. The increase comes from the first term in (\ref{fourfour}).
Again, the maximum value reached lies between one and two times the maximum
for single-time coarse-grainings by the number of balls in one urn. This
behavior is also evident in Figure 4 (though for large $t_1$ the increase
is almost saturated at the initial time).

Increasing the number of times included in each history
is a fine-graining. At a given value of $t_1$, the entropy should
decrease as more times are included. This behavior is
illustrated in Figure 5 for $2R=n_0=30$. This shows the behavior
of one, two, and three time history space entropies as a function of
$t_1$ where $t_2=t_1+1$, and $t_3=t_1+2$.
All the entropies increase
to maximum values on roughly the time scale $R$. Asymptotically from
(\ref{fourfour}), the entropies behave like
\begin{equation}
S_1(t_1)-c
\label{fourten}
\end{equation}
where $S_1$ is the single-time entropy and $c$ is independent of $t_1$ for
the urn model but depends on the number of times and the values of the
time differences.

\centerline{\epsfysize=5.00in \epsfbox{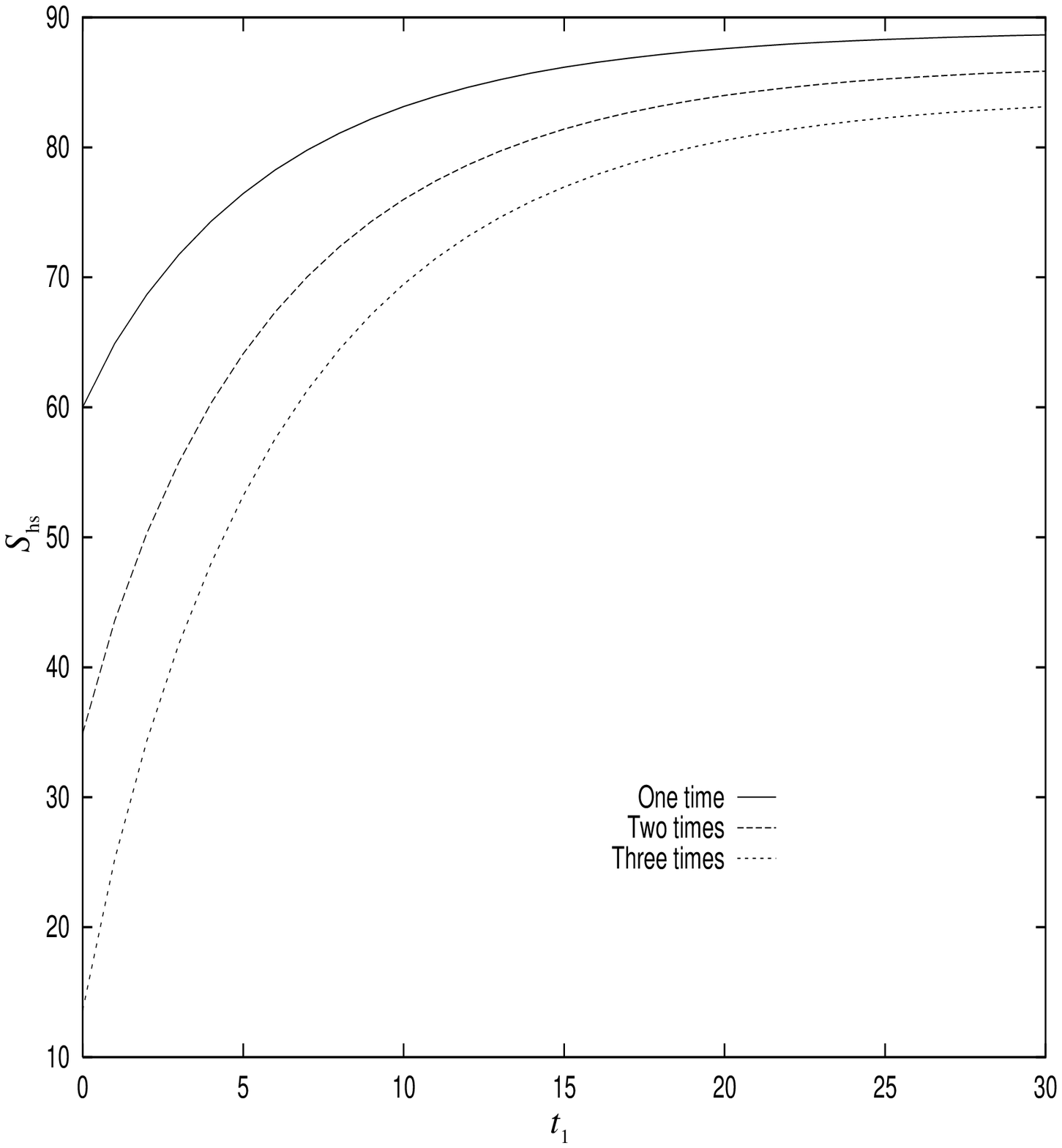}}

\begin{quote} {{\bf Figure 5:}
\sl History space entropy, $S_{hs}$, for
one, two, and three time histories of the Ehrenfest urn model versus
the first specified time $t_1$, in bits.  The times of the two and three time
histories are separated by single timesteps.  The parameters and
initial conditions are the same as in Figure 4.}
\end{quote}

\section{Quantum History Space Entropy}

Isham and Linden posited their family of entropies (\ref{A}) on the basis
of the property that they decrease under fine-graining. We were able to
show that the classical analogs could be derived from a Jaynes
construction for the case $x=1$.  In this section we show that quantum
history space entropy can be similarly derived as a preliminary to 
a more general discussion of its connection with other
entropies.\footnote{The authors have benefited from many discussions
with M. Gell-Mann on this issue.}

Consider a set of decoherent alternative histories $\{c_\alpha\}$, each
history with a probability $p_\alpha$ and represented in history space by a
projector ${\bf P}_\alpha$. Define an entropy functional on history space
operators ${\bf \widetilde W}$ by 
\begin{equation}
{\cal S} ({\bf \widetilde W}) = - \tr\left({\bf \widetilde W}\, \log_2
\, {\bf
\widetilde W}\right)\ .
\label{fiveone}
\end{equation}
Then maximize ${\cal S} ({\bf \widetilde W})$
over all ${\bf \widetilde W}$ for
which (\ref{fiveone}) is real, subject to the condition that
\begin{equation}
\tr\left({\bf P}_\alpha \widetilde {\bf W}\right) = p_\alpha\ .
\label{fivetwo}
\end{equation}
The result is that the maximum is given by
\begin{equation}
\widetilde {\bf W} = \sum_\alpha\, p_\alpha\ \frac{{\bf P}_\alpha}
{\tr\, [{\bf P}_\alpha]}\ ,
\label{fivethree}
\end{equation}
and the entropy is:
\begin{equation}
\label{AAA}
S_{hs}(\{c_\alpha\}) = - \sum_\alpha p_\alpha \log_2 p_\alpha + \sum_\alpha 
p_\alpha \log_2  \tr({\bf P}_\alpha)\ .
\end{equation}
analogous to (\ref{twofifteen}).

The Jaynes construction immediately makes clear why the $x=1$ history space
entropy decrease on fine-graining. There are more conditions constraining
the maximum in (\ref{twotwelve}) in a fine-graining of a set than in the
set itself. The maximum can therefore only be lower. For other values of
$x$ it is sufficient to note that
\begin{equation}
I_x\left(\left\{c_\alpha\right\}\right) =
  S_{hs} \left(\left\{c_\alpha\right\}\right)
  - \tr\left({\bf I}\right)
  + (x-1) \sum_\alpha
  p_\alpha \log_2 \left[\tr \left({\bf P}_\alpha\right)\right]\ .
\label{fivefour}
\end{equation}
This too decreases with fine-graining, as follows from the result for $I_1$
and the convexity of the logarithm.

Thus, history space entropy can be given a unified construction through a
Jaynes procedure both classically and quantum mechanically. What can be
done classically but {\it not} quantum mechanically is to express the
probabilities for {\it all} decoherent histories in the form
\begin{equation}
p_\alpha = \tr\, \left({\bf P}_\alpha {\bf W}\right)
\label{fivefive}
\end{equation}
for one positive operator ${\bf W}$, independent of the set
of alternatives. There is no quantum
mechanical analog of (\ref{twoseven}). Were there one, quantum
mechanics would be equivalent to a classical stochastic theory. It {\it is}
possible to find history space operators ${\bf W}$ which reproduce the
probabilities $p_\alpha$ through (\ref{fivefive}) for any
decoherent set. For example, valid
expressions for the probabilities of decoherent histories like
\begin{equation}
p_\alpha = \tr\, \left(P^n_{\alpha_n} (t_n) \cdots P^1_{\alpha_1}
(t_1)\, \rho\right)
\label{fivesix}
\end{equation}
can be transcribed into history space using the identity \cite{ILS94}
\begin{equation}
\tr_{\cal H} \left(A_1\cdots A_n\right) = \tr_{\otimes^{k_{\cal H}}}
\left[\left(A_1\otimes\cdots\otimes A_n\right)\, S \right]
\label{fiveseven}
\end{equation}
where
\begin{equation}
S\left |v_1\rangle \otimes\cdots\otimes\right | v_k\rangle =
\left |v_k\rangle \otimes |v_1\rangle \otimes\cdots\otimes
\right | v_{k-1}\rangle\ .
\label{fiveeight}
\end{equation}
However, the resulting ${\bf W}$'s are not positive, even when they can be
arranged to be Hermitean.  For this reason, even though quantum analogs
of $S_{dc}(\{c_\alpha\})$ and $S_{ic}(\{c_\alpha\})$ can be defined, the
derivations of the inequalities relating them to $S_{hs}(\{c_\alpha\})$ like 
(\ref{twothirtysix}) do not immediately generalize to quantum
mechanics.

\section{Conclusions}

Information is contained not only in sets of alternatives at a single
moment of time, but more generally in sets of alternative histories
--- sequences of sets of alternatives at a series of times. A variety of
measures of the information in histories are available. In this paper we
have provided a unified construction of all of these through the Jaynes
procedure. It follows from these constructions that these entropies
decrease under fine-graining and increase under coarse-graining. We
illustrated this in a few simple models.

We expect entropies for histories to share other common properties
analogous to the usual second law of thermodynamics. In particular, the
entropy of a set of histories should increase as that set is translated
forward in time away from a low entropy initial condition. We illustrated this with the classical urn model, but
expect it to hold for more realistic dynamical laws, both classically
and quantum mechanically.

General sets of alternative coarse-grained histories will not exhibit
deterministic correlations in time in a classical stochastic theory.
However, sufficiently coarse-grained sets of histories {\it may}
exhibit deterministic behavior.
For example, the unpredictable motion of single atoms yields nearly
deterministic laws for the hydrodynamic variables of pressure,
temperature, and density.  Characterizing the level of
determinism is an interesting question related to the search for
measures of classicality in quantum theory. It is clear from our discussion
that no entropy of histories is a measure of determinism. Entropy is
reduced by fine-graining, and the finest-grained histories are not
deterministic. In quantum theory we can, therefore, not expect an entropy
of of histories, by itself, to be a measure of classicality.

\acknowledgments

We would like to thank Carl Caves and Murray Gell-Mann for useful
discussions. The work of T.~Brun was supported in part by NSF grant
PHY94-07194 and that of J.B.~Hartle by NSF grants PHY95-07065 and
PHY94-07194.

\end{document}